# MULTIVARIANT BRANCHING PREDICTION, REFLECTION, AND RETROSPECTION

**Mark Burgin**

Department of Mathematics
University of California, Los Angeles
Los Angels, CA 90095

**Walter Karplus**

Department of Computer Science
University of California, Los Angeles
Los Angeles, CA 90095

**Damon Liu**

Department of Computer Science
University of California, Los Angeles
Los Angeles, CA 90095

**ABSTRACT**

In branching simulation, a novel approach to simulation presented in this paper, a multiplicity of plausible scenarios are concurrently developed and implemented. In conventional simulations of complex systems, there arise from time to time uncertainties as to which of two or more alternatives are more likely to be pursued by the system being simulated. Under these conditions the simulationist makes a judicious choice of one of these alternatives and embeds this choice in the simulation model. By contrast, in the branching approach, two or more of such alternatives (or branches) are included in the model and implemented for concurrent computer solution.

The theoretical foundations for branching simulation as a computational process are in the domains of alternating Turing machines, molecular computing and E-machines. Branching simulations constitute the development of diagrams of scenarios representing significant, alternative flows of events. Logical means for interpretation and investigation of the branching simulation and prediction are provided by the logical theories of possible worlds, which have been formalized by the construction of logical varieties. Under certain conditions, the branching approach can considerably enhance the efficiency of computer simulations and provide more complete insights into the interpretation of predictions based on simulations.

As an example, the concepts developed in this paper have been applied to a simulation task that plays an important role in radiology - the noninvasive treatment of brain aneurysms.

**Keywords** - *Virtual reality, Decision support systems, Simulation methodology, Health care, Real-time simulation.*

## 1. INTRODUCTION

It is not necessary to prove importance of simulation as a whole and computer simulation, in particular. However, it is also well known that computer simulation can involve and has involved in its practice different fallacies, some of which may be very dangerous. Consequently, we come to the following problems. How better to estimate role of computer simulation? How to make it more useful to society and more adequate to problems that are solved by computer simulation?

In all these aspects, theory, methodology, and technology play the decisive role. However, their roles are different. Methodology as a tool for human activity gives principles and methods to develop, enhance, and improve processes and procedures. Theory transforms these methods and principles into (more or less) exact structures and general procedures. Technology converts what is known in methodology and theory into concrete procedures and algorithms developing various devices for realization of these procedures and algorithms.

Consequently, methodology has to answer how to understand good and bad in human activity and how to achieve better functioning of a system, utilization of a device, or realization of a process.

The basic philosophical assumption for this is that that there are no good and bad things in themselves. What makes this or that thing good or bad is how people use this thing. For example, you can use electricity for bringing light to your house or for killing people. Using telephone, it is possible to help or to cheat people. Some things are simple and their utilization is simple. In such situations, evaluation and decision making is also comparatively simple. However, in the case of complex and sophisticated systems and processes, problems of what is good and what is bad might be very hard. But even when we have answers to these questions, it is far in many cases from being evident how to improve situation. This is just the problem where methodology is necessary because the principal goal of methodology is to teach us how to use things in a better way.

Computer simulation is used for various purposes. One of the most important is prediction. There are cases when it is possible to predict with a great precision. For example, people learned how to foretell sun eclipses long ago. However, for complex systems, the situation is quite different: predictions are, as a rule, very vague and uncertain (Karplus 1992). Consequently, to get some useful information simulation has to give a multiplicity of plausible scenarios for future outcomes. Even though, as it comes to the situations predicted, neither of these scenarios may fit exactly into the actual development of events. At the same time, it is necessary to know what is going to happen further and how to behave in this or that case. So, when the bad correlation of predicted and real events is found, predictions for more distant time become even more unreliable.

To solve this discrepancy, two approaches are usually utilized. The first of them, which is called sequential, involves the full repetition of the simulation, only with other parameters that correspond to the new data. However, in many cases, such simulation demands too much time to be efficient in a critical situation: either there is no time at all for such simulation, or when the results of the new simulation are obtained, the situation again changes drastically. The second approach, which is called selective, suggests comparison of the current situation to the predicted scenarios and the choice of such a scenario that best fits the real situation, i.e., is the closest to this situation in some metric. This approach assumes that the closest, the most relevant scenario provides a good approximation of reality. Sometimes, it is really so, but the question is what to do when this is not the case. An answer is given by branching simulation that synthesizes the advantages of the both approaches eliminating their shortcomings. In other words, branching simulation makes possible to achieve better approximation to reality than in the selective approach, while utilizing less time than demands the sequential approach.

## 2. COMPUTER EXPERIMENTS

If we analyze computer simulation, we can see that it is a computer experiment. At the same time, the first aim of any experiment is to see what will happen if we do something. More distant aims of experimentation are achieved through reasoning and calculations. For example, when experiments are conducted to test some hypothesis or theory.

All experiments are divided into three classes according to their interaction with the system of the experiment. If the experimental actions have no impact either on the system under consideration or on its environment, then it is an experiment without interference, which is usually called observation. If there is no impact of the experimental actions on the system itself but these actions influence environment of the system, then it is a non-disturbing experiment, If there is impact of the experimental actions on the system itself, then it is a disturbing experiment.

In this way, computer simulation of a system is an experiment with a model of a system. As such, it may be considered as an indirect experiment with the system itself. These computer experiments provide some knowledge about the behavior of the system. Properties of this knowledge (such as correctness, adequacy, precision, validity, reliability, etc.) depend on the three main factors: properties of the used model of the system, properties of computer realization of this model, and properties of the input information that is used in computer simulation.

Computer experiments might be very useful. For example, they can give satisfactory results in many cases when it is either impossible to conduct an actual experiment or such experiment might be too expensive, dangerous, etc. One of the most recent examples of successful computer experimentation is nuclear physics.

However, there are also cases when computer experiments cannot give valid results. For example, in simulating the global climate, it is necessary to guess or to assume how much carbon dioxide will be released into the atmosphere over the next several years. This is input data and it depends heavily on unknown factors such as future energy consumption, the introduction of alternative energy sources, the future of nuclear power, etc. Consequently, this information is ill-defined, and many different forecasts are made - some predicting small changes and others dramatic outcomes. So, in this case, we do not have a good prediction even for the direction in which the global climate develops.

## 3. BRANCHING SIMULATION

To understand better the essentials of branching simulation, it is necessary to consider the conventional simulation process as described in (Karplus 1992). A scientist who succeeds in getting a simulation to run on a computer and who has gained confidence that it is running properly is like a child with a new toy. Unlike a scale model or a laboratory setup, a simulation is infinitely flexible. A few simple commands entered from a terminal can effect profound changes in the model and the way it transforms inputs into outputs. Here is the opportunity to ask: "What would happen if … ?" and to get an answer in short order … for a while a scientist feels like a master of the universe.

In a similar way, engineer uses simulation to consider different options of behavior of the device or software she/he is designing. Each simulation produces some feasible scenarios and each scenario is developed from the very beginning many times.

In contrast to this, branching simulation utilizes more efficiently knowledge obtained in previous runs of the simulation model. Repetitive simulation procedures start not from the very beginning but use common parts of computational routes. For example, let us consider simulation of some process from time $T_B$ to time $T_f$ (see Figure 1). In many of such cases, it is desirable to see what happens when at some point $T_1$ situated between $T_B$ and $T_f$ the development of the process changes and these changes have many options. It is possible to repeat the whole simulation with different initial data and track all directions of the process development. However, such an approach is far from being optimal. More efficient, as a rule, is to preserve the results of simulation at the interval $[T_B, T_1]$ and begin new simulation procedures only from the point $T_1$. Thus, we get branching of processes at this point.

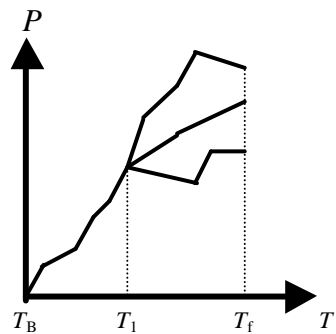

*Figure* 1: Here *T* is time and *P* denotes the process parameters.

Concurrency and parallelism are urgent demands of modern computations in their search for efficiency. The whole process of branching simulation is intrinsically concurrent (and only in some cases completely parallel). But this concurrency differs from the possible or actual parallelism of conventional simulation processes because each of such processes begins from the very beginning without taking in account all knowledge obtainable in a simulation process. However, in many cases when a new simulation process begins, it is possible to take a part of one of the previous simulations as a part of this new simulation and to realize simulation only for what happens after the chosen part ends. Thus, when a person learns how to run, she/he does not begin to do this as if she/he is a new-born individual who knows nothing and can do nothing. On the contrary, the person in question utilizes her/his knowledge and skills in walking because she/he has been already taught how to walk.

Taking this beautiful metaphor, when "a scientist who succeeds in getting a simulation to run on a computer" is compared to a master of universe, we can see an essential difference between this scientist and the real

Master of the Universe. As it is written in the Bible, God did not destroy all human beings and created new ones from the clay when it became evident that the first experiment with people was unsuccessful. On the contrary, God sent the Flood into world and this flood eliminated all people but one family. This family of patriarch Noah gave birth to all new generations of people. That is, a new experiment included some part of the previous way of the mankind. For example, the history of Adam Eve in Eden was not repeated.

Even when a person is being taught relatively different skills like swimming, some knowledge about walking is utilized (especially, when it is the freestyle). However, in this case, there exists much less knowledge about walking that is relevant to swimming than in the case of running. Consequently, it necessitates evaluation of the knowledge obtained in simulation and usage of this evaluation in organization of a directed branching simulation.

The theoretical foundations for branching simulation as a computational process are provided by alternating Turing machines (Chandra et al. 1981), molecular computing (Sipper 1999), and E-machines (Burgin 1980). But contrasting to the goal of these abstract devices which are directed, as a rule to a single goal, branching simulation is aimed at the development of a diagram of scenarios representing significant, alternative flows of events.

## 4. MULTIVARIANT AND BRANCHING PREDICTION

In (Karplus 1992), a qualitative scale for computer simulation is elaborated. According to this scale, it is possible to obtain very good results of mechanical systems and electrical circuits, good results for aircraft control and nuclear reactors. Worse results are related to air pollution and ecological systems. It is almost impossible to predict reliably for social and political systems. Especially, it is almost impossible to predict such rare phenomena as catastrophes because the variables of the used models go out-of-bounds. Consequently, the models that have been constructed and validated over the years are certain to become invalid with the approach to catastrophes. And as a rule, we have no way of validating these models because the catastrophes are too rare. Most have never happened before. Thus, scientists come to the conclusion that catastrophes cannot be predicted scientifically (Karplus 1992). Predictions become very unreliable but they may be still useful serving to alert us to the possibility of a catastrophe.

However, it possible to derive even more profit from computer simulation in the case of catastrophes. To do this, it is necessary to consider a new form of prediction when several possible outcomes are predicted in contrast to the conventional prediction when only one "real" phenomenon or situation is predicted. This kind of prediction, we call the *multivariant* prediction.

This approach is consistent with the physical theory of the multi-world universe (Eastbrook 1998). The essence of this theory is that at each moment (at definite moments) of time the universe is divided into several similar but not identical universes which continue to develop in their own direction dividing on every new step of this development.

Multivariant prediction can also alert us to what may happen but its main goal is to prepare us to these options. To make this idea more precise, it is necessary to explain what does it mean to be prepared to something. Here, we consider only negative situations (e.g., catastrophes) and separate specific levels of preparation.

The first level is to eliminate the negative situation in question as a whole. For example, there is the Y2K problem, routed in the fact that the main body of computer was built and programmed in such a manner that a year is represented only by two digits. After December 31, 1999, this may cause a breakdown of all computer networks. It will be a catastrophe because now human civilization is essentially dependent on computer technology. That is why, great efforts have been made during last years to remedy this situation.

The second level of preparation is to eliminate the negative consequences of the situation in question. For example, winter in some regions of the Earth is very cold. People cannot yet change the climate there. But they can build warm houses and produce warm clothes eliminating in such a way consequences of the frost for themselves. Previously, it has been done empirically. Now in such cases, it is possible to use computer simulation.

The third level is to make the negative consequences of the situation in question as small as possible. For example, it is impossible either to make impossible all car incidents or even to make them safe for drivers and passengers. However, producers of cars are working all the time making cars safer and safer, and computer simulation helps a lot in this process.

With respect to the outcome, all predictions are divided into three classes:
1) *descriptive* prediction demonstrating what will be in future;
2) *prescriptive* prediction representing what is necessary to do in future; and
3) *evaluative* prediction explicating what is possible in future.

The multivariant prediction is evaluative by its nature. However, to prepare us, it is necessary for branching prediction to be also prescriptive. Thus we come to a synthesis of evaluative and prescriptive predictions when it is considered what can happen and what is necessary to do in each of possible situations. We call this type the *conditional* prediction. It has to develop possible scenarios of future events making possible to understand what might happen and what is necessary to do in this or that case. In such a way, it can prepare society to critical situations giving in advance for strategic situations practical knowledge of possible development of events as well as a description of relevant actions in this or that situation. This knowledge has to be stored and kept until a critical situation comes. When time is a critical parameter, in other words, when there is no enough time to make grounded decision on the spot, this stored knowledge may be crucial and even vital. Branching simulation used for prediction is called branching prediction. It makes multivariant prediction more efficient. The salient characteristic of branching prediction that distincts it from the prediction is that different branches of simulation that construct feasible scenarios are developed concurrently based on utilization of previous obtained scenarios and systematic evaluation of the events and routes of the development of events.

Some theoretical results, proved in the theory of algorithms, support these implications. For example, mathematical theory of complexity (Balcazar et al. 1998) makes a distinction between simple and hard problems. Some problems are proved to be very hard because to be solved they demand a lot of time and/or a huge amount of other resources. In theory of complexity, such problems are called NP-complete or PSPACE-complete problems. However, if somebody solves some of these problems in advance and stores the answers in some memory, then finding an answer will take not much time (and other resources). This makes these problems very simple essentially reducing their complexity.

Logical means for interpretation and investigation of the branching simulation and multivariant prediction are provided by the logical theories of possible worlds (Cann 1993), which is formalized by the construction of logical varieties (Burgin 1995). These constructions define semantics for simulation as a whole and for branching simulation, in particular, making possible to apply to them traditional and new logical methodology in researching for simulation processes such problems as realizability, satisfiability, solvability, etc. Many practical problems in programming and computation have formal logical representation, and consequently, they may be treated by logical means that give more founded results than traditional empirical means (Tucker et al. 1994). As an example, we can consider proofs of program correctness. It is only essential to choose adequate representation and appropriate tools.

## 5. BRANCHING REFLECTION

Simulation may be used to represent future, present or past. In the first case, we call it *prediction*. In the second case, it is called *reflection*. The third case is named *retrospection*. Branching simulation approach is useful not only for prediction, but as well for reflection and retrospection. In this section, the main emphasis is on the case of reflection.

It is clear that branching might be very useful for prediction as the future is not uniquely determined and there are different options for which it is difficult to see in advance which of them (if any) will reflect the future situation. At the same time, it looks so as if all events in past and present form a unique sequence. However, branching and multivariance are relevant processes for past and present. For the past, it is relevant because very often it is necessary to reconstruct what happened in the past but we do not have enough knowledge to do this exactly. Consequently, a system of possible event sequences emerges and in their turn, they involve branching as an efficient technique for investigation of different options of what might happened in the past, that is, for a multichoice retrospection. Present emerges from the past through an extensive net of causations. As a result, brainching appears so far a very useful technique for simulation of the present, that is, for reflection.

The reflection techniques are often exploited by real-time simulations that have been applied to different training and education systems. For example, several simulation systems are developed for the medical specialists that allow individual neurosurgeons or radiologists to learn and visualize complex human anatomy and blood flows, interactively explore established clinical procedures in a computer-generated world, without putting patients at risk. For example, all of the radiologists using simulating systems can be expected to have similar objectives and exercise only those aspects of the tasks that are relevant to their needs. The kind of interactions that such users will require from the system are therefore predictable to a considerable extent. There are several ways in which this knowledge of limited number of scenarios can be utilized to maximize the effectiveness of the simulation. To this end, branching might be a promising approach to increase the efficiency of such simulations, making them more adequate to real situations and providing more opportunities to understand the dynamics of the tasks.

However, the prevalent goal in reflective simulation is to present current characteristics and, in many cases, the latest picture of the simulated process. If it is done only once for a definite moment of time, then branching is superfluous. Although, it might happen, the usual case of reflection is repetitive simulation of the same process during some period of time. If the process is changed completely in-between two consecutive simulations, then each simulation cycle has to be realized from the very beginning. However, in many case, only some of all parameters are changing (the *partial parametric stability*) or some parts of the process stay unchanged (the *partial spatial stability*) when the next simulation cycle begins. These features of the process imply that branching can drastically increase efficiency of the simulating system. It is especially essential for interactive real-time systems where time is a critical parameter for simulation.

Two types of stability are related to two types of branching: parametric and spatial.

Technical developments are rapidly improving the computing power and the level of visual realism which provides proper capabilities for the realization of many operative navigation and medical checkups. Such simulations seek to augment or enhance the way medicine is taught and delivered by providing a more effective training and education in a safe, controlled, and consistent fashion. In a simulated environment, inappropriate manipulations are acceptable. However, radiologists are expected to learn from such mistakes before progressing to actual patients. Errors should only be made in simulations and never be repeated in real life experiences.

## 6. VIRTUAL REALITY FOR THE ANEURYSM SURGERY

### 6.1. Medical Problems and Simulation Tools

The methodology of branching simulation in the case of branching reflection has been applied to a simulation task that plays an important role in radiology - the noninvasive treatment of brain aneurysms. Aneurysm surgery remains dangerous because surgeons have limited knowledge of blood flow patterns and complex anatomy of aneurysms. Part of the problem physicians face is to determine if the aneurysm is suitable for a certain surgical technique. For example, if it is too large, in a different location, or oddly shaped, there may be difficulties. Techniques are often considered in an educational environment with its inherent risks to patients. Accordingly, scientific simulation/visualization of time-varying aneurysm anatomy, pressure, and flow at any point in the vascular system is one example of how virtual reality (VR) can help medical specialists roam the complex dataset, evaluate the effects on sealing, blood flow, and pressure of various surgical techniques. Here VR is used for reflection. VR-based systems in medicine represents a paradigm shift in the way that will allow surgeons to evaluate and practice new technologies in a simulator before using them on patients (Chinnock 1994).

At UCLA, researchers are currently building a VR simulation environment to help the neurosurgeon, radiologist, or vascular system specialist plan treatment of brain aneurysms, which are a weakening of the blood vessel walls that can expand like a balloon. Treating aneurysms with implantable coils is designed to seal this expansion to prevent them from bursting (Karplus and Harreld 1994).

The Virtual Aneurysm (VA) system operates on a network that supports client-server paradigm. Flow simulations are computed over time as the heart goes through its pumping cycle. To ensure numerical stability, simulations are computed using small time stepsize such that only a very small fraction of the total data change their values from one step to the next.

Simulations often run for tens or hundreds of hours on high-performance computing machines and periodically generate snapshots of states. The large quantities of simulated data are subsequently stored in archives or databases on disk. After data are off-loaded, they are analyzed and post-processed using VR and scientific visualization techniques to explore the evolving state of the simulated fluid dynamics within the vascular system from local graphics workstations (Liu et al. 1997).

Typical visualizations require a significant amount of I/O bandwidth for accessing data at different time steps when there is not enough memory space for the entire time sequence. The results of data access must be communicated to the graphics workstations, which not only causes significant data movement across slow networks but also interfaces with complex human-computer interactions.

## 6.2. Branching Approach in the VA System

There are two key aspects of computational efficiency of the VA system to which the branching approach can be applied to improve performance.

### 6.2.1. Simulation Aspect

Our Virtual Aneurysm system is based on the numerical solutions of Navier-Stokes equations for the case of three-dimensional time-varying flows. Usually, data describing a mesh geometry is the input, and solution data (velocity, pressure, and so on) at a finite set of points within the computational mesh is the output. The geometry of the mesh is commonly specified by an ordered list of nodes in combination with a list of nodal point coordinates.

Time-varying flow solutions consist of a series of single-time solutions. To ensure numerical stability and desired accuracy of the solutions, it is critical that the time stepsize, and the type and distribution of mesh vertices be controlled. Adding this time and mesh control dramatically increases the dataset size, increasing storage requirements and computational complexity.

The physical system modeled is blood flow in a lateral wall aneurysm with its parent artery. Simulations proceed using small time stepsize and fine grids of mesh elements. Very often, only data values in the regions near or inside the aneurysm have rather noticeable changes, while data values in the other parts of the artery remain static across successive time steps. Furthermore, many simulated data values at a mesh vertex do not differ markedly from data values at its adjacent vertices in the previous time steps.

Consequently, the branching approach implies many advantages to guide effectively the simulation process. Under program control and without user input or intervention, the revised simulation at a vertex can therefore re-use the results of the simulations at its adjacent vertices or in the previous time steps, and continue computing only those data values that actually change. In some instances, this unchanging data make up a large percentage of the solution data, allowing simulation time to be saved by avoiding unnecessary or redundant computations. While this method does incur some overheads in time and memory, it still enhances productivity by offering substantial reductions in overall computational and time costs.

**6.2.2. Visualization Aspect**

Visualization is a simulation of a process with a visual output. The visualization algorithm used to explore CFD data, for example, cutting plane, frequently requires rapid access to subsets of data involves the absolute positions in space. The core technique for loading data into the cutting plane tool is based on a mechanism for returning a value at a given position in space. This position is a 4D space-time point. The value may be a single scalar value like pressure, or a vector quantity like velocity. With a cutting plane, users can view a manageable subset of data and can explore the field by navigating the window through the space.

Flow raw data is generated from a simulation which typically consists of hundreds of time steps worth of information. Due to the spatial and temporal coherence, if two positions are close, the fraction of variation between data is often very small. We can develop a method that exploits the techniques involving the branching approach, especially in combination with this spatial-temporal properties in the simulated data of consecutive time steps or neighboring locations. The delay in cutting plane visualizations can be eliminated by computing and loading only those data values that have changed in subsequent frames. As a result, the amount of computations and I/O bandwidth required for a subsequent frame can be considerably smaller.

## 7. EVALUATION OF BRANCHING

Although branching simulation demonstrates its advantages in applications, theoretical considerations make possible to estimate the gains from branching. Thus, it is possible to separate such cases when branching is beneficial from the cases when it is not worth to implement branching.

Branching as an operation on a simulation process is done at some time $t$ in the functioning of the simulated system **R** when this system is in some state V or when some event B happens in **R**. That is, the branching point is determined by two parameters: system time $t$ and system state V (or system event B). We are going to consider at which points it is worth to apply branching.

**Definition 7.1.** Two trajectories are called I-eqivalent or simply equivalent on the interval $[t_0, t_1]$ if each of them gives the same information as the other one.

**Remark 7.1.** As equivalent trajectories provide the same information, it is not necessary to simulate some trajectoroy if its equivalent one has been already simulated.

**Remark 7.2.** A change in a simulation method or in the problem for which simulation is utilized may transform equivalent trajectories into non-equivalent ones and vice versa.

Let $A$ and $B$ be two events in a system **R.** An event $A$ is, for example, coming of the system **R** to a certain state V. $A\grave{U}t$ denotes that $A$ happens at the time $t$ and Sim $\{\mathbf{R}; A\grave{U}t\}$ denotes simulation of the system **R** with the initial conditions $A\grave{U}t$. **R**/$A\grave{U}t$ denotes the system **R** at time $t$ when the event $A$ has happened. $[\mathbf{R}/A\grave{U}t]$ $(t+\Delta t)$ denotes the results of **R**/$A\grave{U}t$ at the time $t+\Delta t$.

Let $A$ be the initial point of a simulation process (simulation branch) with the time $t_0$ and $t_f$ is the end-time of this process.

**Theorem 7.1.** Branching simulation of **R** gives the same results (demand the same resources, such as time) as the linear simulation of **R** if and only if all trajectories of $[R/A \wedge t_0]$ from $A$ to $B$ are equivalent as well as all trajectories of $[R/A \wedge t_0]$ from $B$ till the time $t_f$ are also equivalent.

Theorem 7.1 gives necessary conditions for branching in a general case. Consequently, Theorem 6.1 also implies when it is worth to make branching. However, to work out exact sufficient conditions, it is necessary to make some additional assumptions.

Let us suppose that:

1) there are no limits for saving the results of simulation in the computer memory;

2) the access time to simulation results is essentially less that the simulation time needed to obtain these results.

**Theorem 7.2.** Branching simulation at an event $B$ with the time $t_1$ saves time of the whole simulation process in any of the following situations:

A) the process $[R/A \wedge t_0] B \wedge t_1$ generates at least two non-equivalent trajectories;

B) all trajectories of the process are equivalent but there are at least two non-equivalent trajectories from $A$ to $B$.

**Remark 7.3.** The conditions for branching in $B$ at time $t_1$ to be reasonable are dependent only on the trajectories going through $B$ at time $t_1$. Other trajectories of $[R/A \wedge t_0]$ ending at time $t_f$ do not influence necessity of branching in $B$. In a general case, there are trajectories of $[R/A \wedge t_0]$ that at time $t_1$ go through another event $C \neq B$.

**Remark 7.4.** In the case B) of the Teorem 4.2, branching at $B$ implies for the simulation from $t_0$ to $t_f$, saving the information obtained from $t_1$ to $t_f$ and simulating only different trajectories from $A$ to $B$.

**Remark 7.5.** The quantity of dissimilar trajectories and thus, the utility of branching may be estimated in advance by theoretical means in many cases. However, more sophisticated evaluation is required to find the actual trade-off for branching.

## 7. CONCLUSION

Thus, a new approach to computer simulation is suggested. It is called the *branching simulation.* Its main peculiarity is that different branches of simulation that construct feasible scenarios are developed concurrently basing on utilization of previously obtained scenarios and systemic evaluation of the events and routes of the development of events.

Branching simulation provides new opportunities for on-line simulation when time is a critical resource. Branching approach makes possible to spare time of computations in several ways. First, excessive computational branches are determined and eliminated. Excessive computational branches appear due to the fact that some regions of the considered vessels do not change in the process of aneurysm treatment. Second, branching simulation provides conditions for elimination of common subsequences of computational sequences.

Such situations, for example, happen rather frequently in a simulation task that plays an important role in radiology - the noninvasive treatment of brain aneurysms. To cure an aneurysm properly, it is necessary to know fluid characteristics in a blood vessel with the aneurysm. These parameters are provided by the simulation system developed at UCLA. The results of simulation are displayed on a screen to help the doctor who is operating the aneurysm. When the simulation and display programs are used in the interactive mode, time becomes a critical parameter.

A new direction in computing that is beneficial for branching simulation is cellular computing or its particular case, molecular computing (Adleman 1994; Sipper 1999). As research shows (Sipper 1999), the properties of cellular computing models are flexible and can be tailored to specific tasks. However, it is not an all-encompassing general-purpose approach but a methodology which can excel being applied to the branching simulation.

Encapsulating benefits of branching, computer simulation as tool for prediction can be useful, at least, in three aspects. First, it can find what can happen in future. It might be not a complete picture of all possible critical events but it can help not to overlook some essential threats and opportunities. Second, computer simulation can provide for evaluation of different situations and actions. Third, it can prepare society to critical situations giving in advance practical knowledge not only what may happen but also what to do in this or that situation. When time is a critical parameter, in other words, when there is no enough time to make grounded decision on the spot, this stored knowledge may save not only a lot of resources but a great quantity of lives.